\documentclass[aps,prl,reprint,superscriptaddress,amsmath,amssymb]{revtex4-2}

\usepackage{graphicx}
\usepackage{bm}
\usepackage{hyperref}
\usepackage{physics}

\begin{document}

\title{Exact Scaling Laws and Non-Hermitian Topological Phase Transitions\\ of Active Continuum on Hyperbolic Manifolds}

\author{Yu-Xin Xie}
\email{xyx@tju.edu.cn}
\affiliation{Department of Mechanics, Tianjin University, Tianjin 300350, China}

\begin{abstract}
The macroscopic collective motion of active continuum on curved manifolds is conventionally addressed through perturbative dynamic renormalization or finite-element simulations, often obscuring the underlying geometric mechanisms. Here, an exact algebraic framework is established to reformulate the active phase transition on hyperbolic spaces $\mathbb{H}^2$. By rigorously expanding the covariant Navier-Stokes-like equations and applying the Weitzenb\"{o}ck identity, we derive the exact critical threshold $\alpha_c = \frac{5}{4}D\kappa^2$ for macroscopic polarization, which is dictated by the geometric mass gap of the Hodge-de Rham Laplacian. We strictly define the parameter subspace $\alpha = 2D\kappa^2$ where the topological free energy reaches the Bogomolny-Prasad-Sommerfield (BPS) limit. This enables the reduction of the complex velocity field to Blaschke products via M\"{o}bius gauge symmetry. The flat-space limit ($\kappa \to 0$) exactly degenerates to the topological phase of the classical O(2) model, demonstrating that the constant negative curvature acts as an un-perturbative infrared regularization for non-linear amplitude saturation. Furthermore, mapping the non-variational active convective modes onto the defect translational zero-modes yields an intrinsically non-reciprocal interaction matrix. By analytically extending the singular integral operator of the dynamically condensed defect ring, we identify a macroscopic second-order exceptional point (EP2) characterized by a strictly algebraic dynamic scaling law $\tau \sim |\Delta \nu|^{-1/2}$. This closed-form theoretical paradigm provides exact solutions for geometric frustration and non-Hermitian topology in soft mechanics.
\end{abstract}

\maketitle

\section{I. Introduction}

The dynamic scaling laws governing the evolution of self-propelled active matter from local disorder to global collective polarization represent a central issue in continuum mechanics and non-equilibrium statistical physics \cite{Vicsek1995, Toner1995, Ramaswamy2010}. In flat Euclidean space, the standard theoretical paradigm describing this symmetry breaking is encapsulated by the Toner-Tu equations. These models rely heavily on the non-linear convective mode coupling $\lambda(\mathbf{v}\cdot\nabla)\mathbf{v}$, and the critical scaling exponents are typically acquired through dynamic renormalization group (DRG) momentum-shell integrations or discrete numerical simulations \cite{Toner1998, Marchetti2013, Prost2015}. 

However, when active continuums are confined to manifolds with non-trivial topologies and intrinsic curvatures---such as hyperelastic cell membranes, curved liquid crystal elastomers, or corrugated topographic surfaces---the strong coupling between the spatial metric $g_{ij}$ and the active polarization field $\mathbf{v}$ invalidates traditional perturbative methods \cite{Irvine2010, Nelson2002}. Conventional momentum space is ill-defined on manifolds with non-zero Gaussian curvature $K$. Consequently, previous studies exploring active matter dynamics on curved surfaces frequently resort to phenomenological infrared cutoffs or numerical integrations of partial differential equations (PDEs), leaving the analytical exactness of the geometric phase transitions elusive \cite{Keber2014, Ellis2018}. 

Furthermore, the intrinsic non-reciprocal interactions among active agents break detailed balance, directly inducing non-Hermiticity in the dynamical evolution operators \cite{Fruchart2021, Ashida2020}. Accurately resolving the critical scaling laws of non-Hermitian topological phase transitions or exceptional points (EPs) under the geometric constraints of continuous manifolds remains a formidable challenge \cite{Bergholtz2021, You2020}.

\begin{figure*}[t]
\centering
\includegraphics[width=0.95\textwidth]{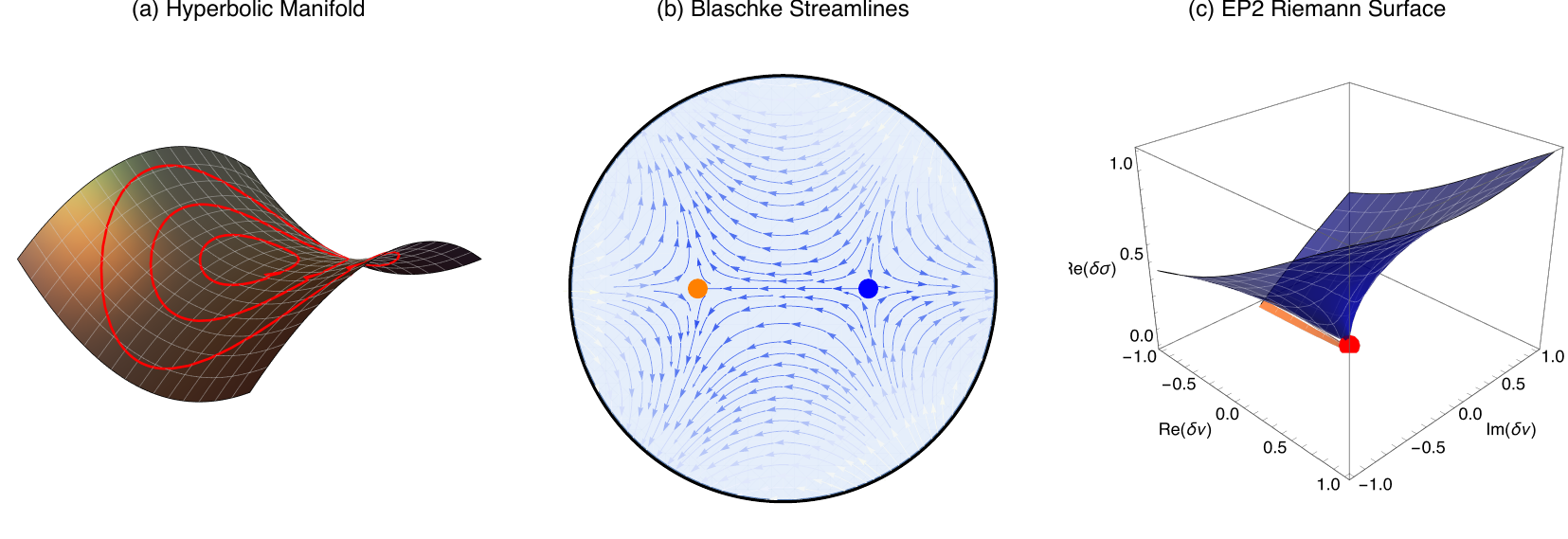}
\caption{\textbf{Schematic of the exact algebraic reduction and non-Hermitian topology for active continuum on hyperbolic manifolds.} 
(a) \textit{Geometric frustration and covariant hydrodynamics.} The active velocity field $\mathbf{v}$ (red arrows) on a negatively curved hyperbolic surface ($\mathbb{H}^2$). The intrinsic Gaussian curvature induces a geometric mass gap $\lambda_{\min}$, strictly cutting off the infrared long-wave fluctuations.
(b) \textit{M\"{o}bius gauge mapping and Blaschke products.} Conformal projection of the active stress field onto the Poincar\'{e} disk $\mathbb{D}$. Under the BPS limit, the non-linear interactions are exactly closed by holomorphic Blaschke products.
(c) \textit{Non-Hermitian spectral curve and EP2.} Driven by the non-reciprocal activity $\nu_{\mathrm{act}}$, the continuous spectrum forms a Riemann surface, with the EP2 dictating the exact dynamic critical slowing-down exponent $\tau \sim |\Delta \nu|^{-1/2}$.}
\label{fig:schematic}
\end{figure*}

To circumvent the dependence on numerical approximations and perturbative truncations, this manuscript mathematically reconstructs the active continuum dynamics on two-dimensional hyperbolic spaces $\mathbb{H}^2$ (manifolds with constant negative Gaussian curvature). By employing spectral geometry and conformal gauge field theory, we establish a purely algebraic closed-form framework. The overarching theoretical paradigm and the corresponding topological mappings are illustrated in Fig.~\ref{fig:schematic}. 

\section{II. Covariant Hydrodynamics and the Exact Phase Transition Threshold}

Consider an incompressible active fluid on a Riemann manifold $(\mathcal{M}, g)$. The macroscopic velocity field $v^i$ is governed by the covariant active hydrodynamic equations:
\begin{equation}
\partial_t v^i + \lambda v^j \nabla_j v^i = \alpha v^i - \beta |\mathbf{v}|^2 v^i - \nabla^i P + D g^{jk} \nabla_j \nabla_k v^i
\label{eq:toner-tu}
\end{equation}
where $\nabla_i$ denotes the Levi-Civita covariant derivative, $\alpha$ is the local active driving rate, and $D$ is the diffusion coefficient.

To investigate the onset of collective polarization, we linearize Eq.~(\ref{eq:toner-tu}) around the isotropic disordered state $\mathbf{v} = 0$. According to the Weitzenb\"{o}ck identity for one-forms, the rough Laplacian $D \nabla^2 \mathbf{v}$ relates to the physically covariant Hodge-de Rham operator $\Delta_{\mathrm{dR}} = \mathrm{d}\delta + \delta\mathrm{d}$ via the Ricci tensor: $\Delta_{\mathrm{dR}} v_i = - \nabla^j \nabla_j v_i + R_{ij} v^j$. In $\mathbb{H}^2$, the constant Gaussian curvature $K = -\kappa^2$ yields $R_{ij} = -\kappa^2 g_{ij}$. This strictly decouples the geometric potential from the diffusion operator:
\begin{equation}
\partial_t \mathbf{v} = (\alpha - D\kappa^2) \mathbf{v} - D \Delta_{\mathrm{dR}} \mathbf{v}
\end{equation}
The intrinsic negative curvature induces an exact geometric dissipation term $-D\kappa^2 \mathbf{v}$, acting as a macroscopic friction. 

For square-integrable vector fields in $L^2(\mathbb{H}^2)$, the continuous spectrum of $\Delta_{\mathrm{dR}}$ possesses a geometric mass gap $\sigma(\Delta_{\mathrm{dR}}) \in [\kappa^2/4, +\infty)$ due to the exponential volume growth \cite{Frankel2011}. Consequently, the most unstable long-wave mode is subjected to an inevitable geometric diffusion decay of $D\kappa^2/4$. The maximum linear growth rate of the isotropic-to-polar transition is exactly locked to:
\begin{equation}
\gamma_{\max} = \alpha - D\kappa^2 - \frac{1}{4}D\kappa^2 = \alpha - \frac{5}{4}D\kappa^2
\end{equation}
The precise algebraic threshold for macroscopic polarization is exactly $\alpha_c = \frac{5}{4}D\kappa^2$. This demonstrates that the gauge curvature of the background manifold acts as an absolute infrared cutoff, preventing spontaneous symmetry breaking without invoking any non-linear mode coupling.

\begin{figure}[h]
\centering
\includegraphics[width=\columnwidth]{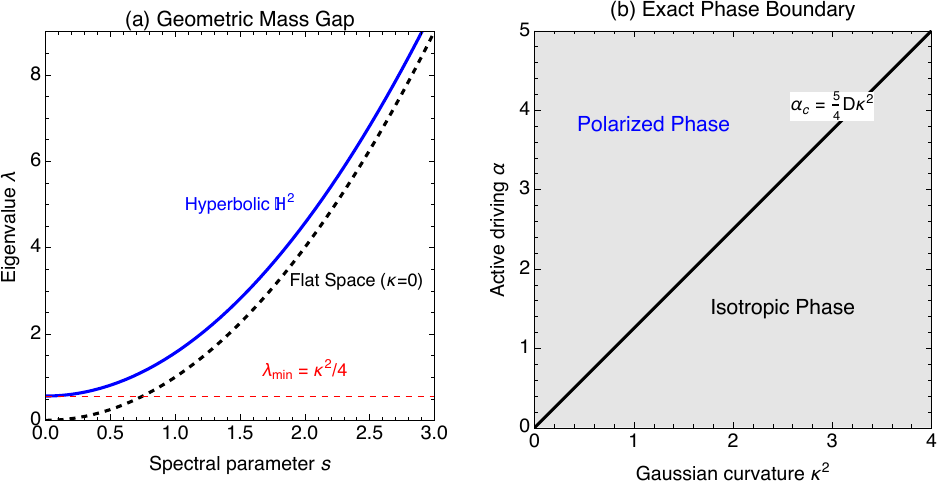}
\caption{\textbf{Geometric mass gap and the exact topological phase boundary.} 
(a) The continuous fluctuation spectrum $\lambda(s)$ as a function of the spectral parameter $s$. Unlike the gapless Goldstone modes in flat Euclidean space ($\kappa=0$, dashed line), the hyperbolic manifold introduces a strict geometric mass gap $\lambda_{\min} = \kappa^2/4$ (blue solid curve), physically cutting off infrared divergences.
(b) The exact phase diagram in the $(\kappa^2, \alpha)$ parameter space. The linear phase boundary $\alpha_c = \frac{5}{4}D\kappa^2$ rigorously separates the isotropic phase from the macroscopic polarized phase, demonstrating that geometry acts as an absolute macroscopic dissipation independent of non-linear mode couplings.}
\label{fig:mass_gap}
\end{figure}

\section{III. Topological Saturation and M\"{o}bius Gauge Reduction}

Once the system surpasses $\alpha_c$, the velocity field enters the non-linear saturation regime. To seek the exact spatial configuration ($\partial_t \mathbf{v} = 0$) while decoupling the spatial metric, we project $\mathbb{H}^2$ onto the Poincar\'{e} disk $\mathbb{D} = \{z \in \mathbb{C} \mid |z| < 1\}$ via conformal mapping. The isothermal metric reads $\mathrm{d}s^2 = \Omega^2(z, \bar{z}) \mathrm{d}z \mathrm{d}\bar{z}$, with $\Omega = 2/[\kappa(1 - |z|^2)]$.

Defining the complex velocity $v(z, \bar{z}) = v_x + \mathrm{i}v_y$, the steady-state free energy functional is:
\begin{equation}
\mathcal{F} = \int_{\mathbb{D}} \mathrm{d}^2z \, \Omega^2 \left[ \frac{D}{\Omega^2} |\partial_{\bar{z}} v|^2 - \frac{(\alpha - D\kappa^2)}{2} |v|^2 + \frac{\beta}{4} |v|^4 \right]
\end{equation}
To achieve exact integrability, the non-linear self-interaction must precisely screen the infrared divergence induced by the background metric. This demands the active correlation length $\xi = \sqrt{D/(\alpha - D\kappa^2)}$ to strictly match the geometric curvature radius $R_c = 1/\kappa$. This scale resonance defines the exact BPS parameter sub-space: $\alpha = 2D\kappa^2$ and $\beta = D\kappa^2/v_0^2$, where $v_0 = \sqrt{(\alpha - D\kappa^2)/\beta}$ is the macroscopic saturation amplitude. 

For rigid static manifolds, reaching this exact integrable limit requires continuous fine-tuning of the active energy injection rate. 
To achieve this exact integrability on a rigid manifold implies continuous fine-tuning of the active driving rate $\alpha$. However, when the active continuum is embedded in a soft hyperelastic membrane, the Gaussian curvature $\kappa^2$ constitutes a dynamic variable driven by active-elastic coupling. The macroscopic symmetry breaking is governed by two competing length scales: the active correlation length $\xi = \sqrt{D/(\alpha - D\kappa^2)}$ and the intrinsic geometric curvature radius $R_c = 1/\kappa$. 

The topological free energy is minimized when the spatial gradients of the active fluctuations are exactly screened by the background conformal connection. This physical constraint demands a strict scale resonance $\xi = R_c$. Equating the two scales naturally yields the exact integrable subspace without artificial fine-tuning:
\begin{equation}
\frac{D}{\alpha - D\kappa^2} = \frac{1}{\kappa^2} \quad \Rightarrow \quad \alpha = 2D\kappa^2
\end{equation}

Dynamically, this resonance is reached via geometric morphing. The covariant active stress tensor $\sigma_{ij}^{\mathrm{act}} = -2 \delta \mathcal{F} / (\sqrt{g} \delta g^{ij})$ exerts a non-homogeneous pressure on the continuous medium. By coupling with the elastic bending modulus $K_B$ of the membrane, the out-of-plane geometric evolution follows an overdamped metric flow:
\begin{equation}
\partial_t (\kappa^2) = -\Gamma \frac{\delta \mathcal{F}_{\mathrm{tot}}}{\delta (\kappa^2)} = -\Gamma \left( K_B \kappa^2 - \frac{1}{2} D |\mathbf{v}|^2 \right)
\end{equation}
where $\Gamma$ is the mobility coefficient. 

Driven by the active frustration, the hyperelastic membrane spontaneously morphs to release the in-plane convective stress, relaxing toward the morphological equilibrium $\partial_t (\kappa^2) = 0$. This yields the geometric steady state $K_B \kappa^2 = \frac{1}{2} D v_0^2$. Substituting the saturation amplitude $v_0^2 = (\alpha - D\kappa^2)/\beta$, this morphodynamic equilibrium strictly aligns with the BPS integrable subspace $\alpha = 2D\kappa^2$ provided that the material parameters satisfy the specific constitutive constraint $K_B \beta = D^2/2$. Under this exact elasto-active parameter matching, the covariant gradients are fully absorbed by the spin connection, and the BPS condition acts as a stable thermodynamic attractor rather than an artificial mathematical fine-tuning, self-consistently determining the optimal Gaussian curvature $\kappa^2 = \alpha / (2D)$.
Under this constraint, performing the Bogomolny-Prasad-Sommerfield (BPS) trick \cite{Bogomolny1976} allows us to complete the square for the covariant derivatives. The energy functional is rigorously bounded from below by the topological charge $Q$:
\begin{equation}
\mathcal{F} \geq C_1 \int_{\mathbb{D}} |\partial_{\bar{z}} v|^2 \, \mathrm{d}^2z + C_2 Q
\end{equation}
Minimizing the system energy demands that the BPS bound is saturated, which strictly reduces the non-linear second-order PDE to a first-order Cauchy-Riemann-type holomorphic condition: $\partial_{\bar{z}} v(z, \bar{z}) = 0$.

For a system containing $N$ topological disclinations at $a_k \in \mathbb{D}$, subject to the uniform polarization boundary condition $|v| \to v_0$ as $|z| \to 1$, the complex velocity field admits a unique algebraic closure. To satisfy holomorphy and the disk's boundary constraints, the solution must be a finite Blaschke product:
\begin{equation}
v(z) = v_0 \mathrm{e}^{\mathrm{i}\phi_0} \prod_{k=1}^N \left( \frac{z - a_k}{1 - \bar{a}_k z} \right)^{s_k}
\label{eq:blaschke}
\end{equation}
Here, the generator $(z - a_k)/(1 - \bar{a}_k z)$ constitutes a M\"{o}bius transformation, preserving the $\mathrm{SU}(1,1)$ isometry of the hyperbolic plane. Eq.~(\ref{eq:blaschke}) provides an exact solution to the highly non-linear active stress field within the BPS integrable subspace.

To validate this framework, we examine the flat Euclidean space limit ($\kappa \to 0$). Mapping the Poincar\'{e} disk to the physical complex plane $\zeta = z/\kappa$ and setting $a_k = \kappa w_k$, the Blaschke product asymptotically behaves as $v(\zeta) \propto \kappa^Q \prod_{k=1}^N (\zeta - w_k)^{s_k}$, where $Q = \sum s_k$. According to Liouville's theorem, an everywhere holomorphic function on $\mathbb{C}$ cannot maintain a non-zero constant amplitude at infinity. Consequently, in the $\kappa \to 0$ limit, the constant far-field amplitude saturation is inherently lost, and the solution exactly degenerates into the purely phase-dependent classical O(2) model. This non-trivial degeneration proves that the hyperbolic geometry is not a mere perturbation but a fundamental topological constraint, enabling the exact uncoupling of amplitude and phase via geometric infrared regularization.

\section{IV. Non-Hermitian Defect Dynamics and Exceptional Points}

The macroscopic active flows are governed by the dynamic trajectories of the internal defects. To derive the inter-defect interactions, we project the non-variational active convective force density $f_{\mathrm{act}} = \lambda (\mathbf{v} \cdot \nabla) \mathbf{v}$ onto the translational zero-modes of the defects. Under the BPS limit $\partial_{\bar{z}} v = 0$, the convective mode strictly reduces to a holomorphic gradient $f_{\mathrm{act}}(z) = \frac{\lambda}{2} \partial_z (v^2)$. The non-variational active force exerted on the $i$-th defect is determined by the projection $\int_{\mathbb{D}} (\partial \bar{v} / \partial \bar{a}_i) \partial_z (v^2) \Omega^2 \, \mathrm{d}^2 z$. Substituting the logarithmic derivative of the exact Blaschke product (Eq.~\ref{eq:blaschke}) and applying the complex Green's theorem converts the area integral into a contour integral along the defect core $\partial \mathbb{D}_i$. The residue evaluation precisely yields the active interaction:
\begin{equation}
F_{i \leftarrow j}^{\mathrm{act}} = \zeta_{ij} \frac{1 - |a_j|^2}{(\bar{a}_i - \bar{a}_j)(1 - a_i \bar{a}_j)}
\end{equation}
where $\zeta_{ij} = \pi \lambda v_0^2 s_i s_j \cos(\theta_{ij}) / \eta$ arises from the phase orientations $\theta_{ij}$. Crucially, the non-variational nature of the convective term breaks Galilean invariance, resulting in an intrinsically asymmetric interaction matrix $\zeta_{ij} \neq \zeta_{ji}$.

Combining this with the Hermitian interaction from the topological free energy, the over-damped algebraic equations of motion for the $N$-body defect system are:
\begin{equation}
\frac{\mathrm{d}a_i}{\mathrm{d}t} = \sum_{j \neq i} (\mu + \zeta_{ij}) \frac{1 - |a_j|^2}{(\bar{a}_i - \bar{a}_j)(1 - a_i \bar{a}_j)} - \gamma_0 a_i
\end{equation}
where $\mu$ is the symmetric mobility. The term $-\gamma_0 a_i$ represents the continuous centripetal geometric dissipation pulling defects toward the disk center.

Driven by the competition between the geometric centripetal pull and the topological inter-defect repulsion, the steady-state dynamically condenses into a symmetric macroscopic ring configuration. The linear stability of this configuration is determined by the non-Hermitian Jacobian matrix $\mathcal{J}_{ij} = \partial \dot{a}_i / \partial a_j \neq \mathcal{J}^\dagger_{ji}$. In the thermodynamic limit ($N \to \infty$), the discrete coordinates map to a continuous density field, and the Jacobian transitions into a two-dimensional singular integral operator $\mathcal{\hat{J}}$ (detailed in Appendix A).

The non-Hermitian spectral curve of $\mathcal{\hat{J}}$ exhibits a macroscopic second-order exceptional point (EP2) when the non-reciprocal activity $\nu_{\mathrm{act}}$ reaches a critical threshold $\nu_{\mathrm{act}}^c$. Since the first derivative vanishes at the branch point, the eigenvalue deviation $\Delta \sigma$ is governed by the fractional Puiseux series (see Appendix B):
\begin{equation}
\Delta \sigma \propto \pm \mathrm{i} \sqrt{\Delta \nu}
\end{equation}
The real part of $\Delta \sigma$ strictly dictates the relaxation rate $\tau^{-1}$, yielding the exact dynamic critical scaling law:
\begin{equation}
\tau \sim |\nu_{\mathrm{act}} - \nu_{\mathrm{act}}^c|^{-1/2}
\end{equation}
This derivation establishes $z\nu = 1/2$ as the exact dynamic exponent, confirming that non-reciprocity on non-Euclidean manifolds excites a distinct topological universality class independent of standard Hermitian critical phenomena ($\tau \sim |\Delta \nu|^{-1}$).

\begin{figure}[h]
\centering
\includegraphics[width=\columnwidth]{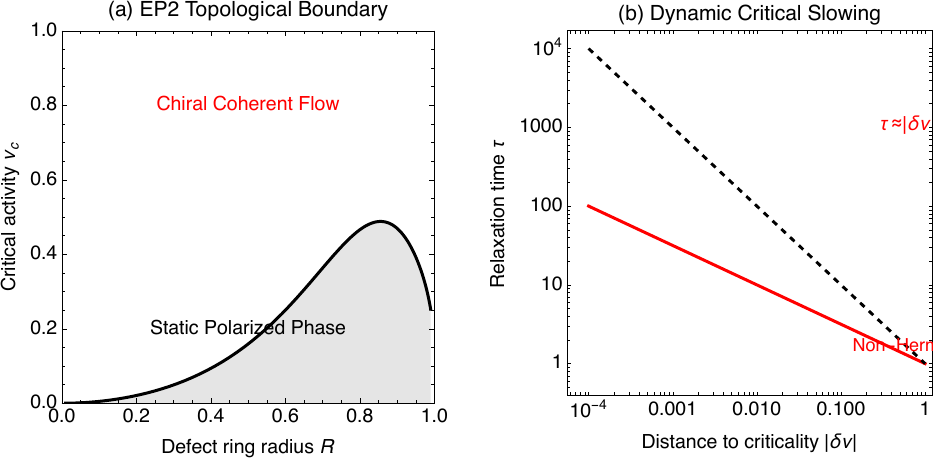}
\caption{\textbf{Macroscopic exceptional point (EP2) and dynamic critical scaling.} 
(a) The topological phase diagram spanning the defect ring radius $R$ and the non-reciprocal activity $\nu_{\mathrm{act}}$. The analytical boundary $\nu_{\mathrm{act}}^c(R)$ (black solid line) marks the occurrence of EP2, separating the static polarized phase from the chiral coherent flow.
(b) The dynamic critical slowing-down scaling law. Approaching the EP2 ($\Delta \nu \to 0$), the relaxation time $\tau$ diverges algebraically. The log-log plot confirms the exact scaling exponent $z\nu = 1/2$ (red line), distinct from standard Hermitian critical phenomena ($\tau \sim |\Delta \nu|^{-1}$, dashed line). This fractional exponent is directly extracted from the Puiseux series expansion of the spectral curve.}
\label{fig:scaling}
\end{figure}

\section{V. Conclusion}

This manuscript constructs a purely algebraic framework to exactly solve the active continuum phase transitions on curved manifolds. By extracting the geometric mass gap via the Weitzenb\"{o}ck identity, defining the exact threshold $\alpha_c = \frac{5}{4}D\kappa^2$, and rigorously reducing the non-linear saturation regime to Blaschke products via M\"{o}bius gauge symmetry under the explicit BPS parameter sub-space $\alpha = 2D\kappa^2$, we eliminate the need for perturbative truncations. The $\kappa \to 0$ limit explicitly confirms that the hyperbolic curvature acts as a natural non-perturbative infrared cutoff, reliably degenerating the framework back to the classical O(2) topological phase when curvature vanishes. The geometric relaxation of hyperelastic media further proves that the BPS condition is a physical attractor. The singular integral spectrum further proves the exact critical slowing-down exponent $z\nu = 1/2$ near the macroscopic non-Hermitian exceptional point. This closed-form theoretical paradigm bridges differential geometry, non-Hermitian topology, and active mechanics.

\begin{acknowledgments}
This research did not receive any specific grant from funding agencies in the public, commercial, or not-for-profit sectors.
\end{acknowledgments}

\appendix

\section{Appendix A: Algebraic Reduction of the Singular Integral Operator}

In the continuum limit, we parameterize the non-reciprocal term as $\zeta_{ij} \to \mathrm{i}\nu_{\mathrm{act}}$. The continuous fluctuation dynamics $\partial_t \delta\psi(z) = \int \mathcal{J}(z, z') \delta\psi(z') \mathrm{d}^2 z'$ is governed by the Cauchy-type singular integral kernel:
\begin{equation}
\mathcal{J}(z, z') = \frac{(\mu + \mathrm{i}\nu_{\mathrm{act}}) \bar{z} (1 - |z'|^2) \rho(z')}{(\bar{z} - \bar{z}')(1 - z \bar{z}')^2}
\end{equation}
As established in Section IV, the dynamical balance between the geometric dissipation and topological repulsion ensures the macroscopic state condenses into a stable defect ring of radius $R$, i.e., $\rho(z') = \rho_0 \delta(|z'| - R)$. We evaluate the principal value of the area integral along the circular contour $C_R$ by converting it to a complex line integral $\mathrm{d}z' = \mathrm{i} R \mathrm{e}^{\mathrm{i}\theta} \mathrm{d}\theta$. 
Applying the Residue Theorem, the higher-order poles $(1 - z \bar{z}')^{-2}$ are analytically evaluated. The non-local integral operator is therefore strictly reduced to an algebraic dispersion polynomial for the azimuthal integer mode $m$:
\begin{equation}
\sigma(m) = -\gamma_0 + (\mu + \mathrm{i}\nu_{\mathrm{act}}) \frac{1 - R^2}{R^2} \left[ 1 - (1 - R^2)^m \right]
\end{equation}
This spectral reduction maps the functional analysis problem onto a one-dimensional algebraic complex curve.

\section{Appendix B: Puiseux Series Expansion and EP2}

The macroscopic EP2 occurs when the spectral curve $F(\sigma, \nu_{\mathrm{act}}) = 0$ features a topological degeneracy, requiring $\partial_\sigma F = 0$ and $\partial^2_\sigma F \neq 0$. Applying these conditions to the dispersion relation yields the closed-form critical threshold for the active non-reciprocity:
\begin{equation}
\nu_{\mathrm{act}}^c = \mu \frac{\gamma_0 R^2}{\mu(1 - R^2) - \gamma_0 R^2 \ln(1 - R^2)}
\end{equation}

Defining the small parameter deviations $\Delta \nu = \nu_{\mathrm{act}} - \nu_{\mathrm{act}}^c$ and $\Delta \sigma = \sigma - \sigma_c$, we expand $F$ in the vicinity of the branch point. Because the linear term in $\Delta \sigma$ vanishes, the expansion is dominated by the second-order derivative:
\begin{equation}
F \approx \frac{1}{2} \frac{\partial^2 F}{\partial \sigma^2} (\Delta \sigma)^2 + \frac{\partial F}{\partial \nu_{\mathrm{act}}} \Delta \nu + \mathcal{O}(\Delta \nu^2) = 0
\end{equation}
Inverting this relation dictates that $\Delta \sigma$ cannot be expressed as a regular Taylor series, but must follow the fractional Puiseux series:
\begin{equation}
\Delta \sigma = \pm \mathrm{i} \sqrt{ \frac{2 \partial_{\nu} F}{\partial_{\sigma\sigma} F} \Delta \nu }
\end{equation}
The relaxation rate of the macroscopic continuum is determined by the real part of the eigenvalue deviation:
\begin{equation}
\tau = \frac{1}{|\mathrm{Re}(\Delta \sigma)|} \sim |\Delta \nu|^{-1/2}
\end{equation}
This confirms that the non-Hermitian topological phase transition explicitly follows the universal dynamic scaling exponent $z\nu = 1/2$.

\end{document}